\documentclass[letterpaper, 10pt,conference]{ieeeconf}     
\IEEEoverridecommandlockouts                          
\usepackage[english]{babel}
\usepackage{amsmath, amssymb, amsfonts, amsthm, bm}
\usepackage{mathrsfs}
\usepackage[mathcal]{euscript}
\usepackage{enumerate}
\usepackage{mathbbol}
\usepackage{float}
\usepackage{color,graphicx,epstopdf}
\usepackage{dsfont}
\usepackage{hyperref}
\usepackage{caption}
\usepackage{subcaption}
\usepackage{bm}
\usepackage{cite} 
\usepackage{tikz-cd}
\usepackage{mathtools}
\usepackage{lipsum}
\usepackage{algorithm} 
\usepackage{algorithmic} 

\newcommand*\mcapinn[2]{\vcenter{\hbox{$\mathsurround=0pt
			\ifx\displaystyle#1\textstyle\else#1\fi\bigcap$}}}

\newcommand*\mcupinn[2]{\vcenter{\hbox{$\mathsurround=0pt
			\ifx\displaystyle#1\textstyle\else#1\fi\bigcup$}}}

\def\range{\textnormal{range}}

\newtheorem{theorem}{Theorem}
\newtheorem{definition}{Definition}

\newtheorem{lemma}{Lemma}
\newtheorem{remark}{Remark}

\newtheorem{assumption}{Assumption}

\newtheorem{experiment}{Experiment}

\newcommand{\rmV}{\mathrm{V}}
\newcommand{\rmG}{\mathrm{G}}
\newcommand{\rmE}{\mathrm{E}}
\newcommand{\rmN}{\mathrm{N}}
\newcommand{\rmO}{\mathrm{O}}
\newcommand{\bfG}{\mathbf{G}}
\newcommand{\bfg}{\mathbf{g}}
\newcommand{\bfx}{\mathbf{x}}
\newcommand{\bfd}{\mathbf{d}}
\newcommand{\bfb}{\mathbf{b}}
\newcommand{\bfy}{\mathbf{y}}
\newcommand{\bff}{\mathbf{f}}
\newcommand{\bfq}{\mathbf{q}}
\newcommand{\bfv}{\mathbf{v}}

\newcommand{\bfD}{\mathbf{D}}

\newcommand{\mcA}{\mathcal{A}}
\newcommand{\mbR}{\mathbb{R}}
\newcommand{\phib}{\bm{\phi}}

\newcommand{\betab}{\bm{\beta}}

\title{\LARGE \bf Differentially Private Games via Payoff Perturbation
}
\author{Yijun Chen, Guodong Shi
	\thanks{Y. Chen and G. Shi are with the  Australian Center for Field Robotics, The University of Sydney, NSW, Australia. (E-mail: yijun.chen@sydney.edu.au; guodong.shi@sydney.edu.au)}
} 

\begin{document}

	\maketitle
	\thispagestyle{empty}
	\pagestyle{empty}

	\begin{abstract}
	In this paper, we study network games where players are involved in information aggregation processes subject to the differential privacy requirement for players’ payoff functions. We propose a Laplace linear-quadratic functional perturbation (LLQFP) mechanism, which perturbs players' payoff functions with linear-quadratic functions whose coefficients are produced from truncated Laplace distributions. For monotone games, we show that the LLQFP mechanism maintains the concavity property of the perturbed payoff functions, and produces a perturbed NE whose distance from the original NE is bounded and adjustable by Laplace parameter tuning. We focus on linear-quadratic games, which is a fundamental type of network games with players' payoffs being linear-quadratic functions, and derive explicit conditions on how the LLQFP mechanism ensures differential privacy with a given privacy budget. Lastly, numerical examples are provided for the verification of the advantages of the LLQFP mechanism.
	\end{abstract} 
	
	\section{Introduction} 	
	
	Games on networks has gained
	increased traction in recent years. It has been applied in a variety of fields such as online E-commerce in social networks\cite{leng2021privacy}, route planning in transportation networks\cite{bianco2016game}, and resource allocations in wireless communication networks\cite{niyato2008competitive}. There are typically three information aggregation processes in games for players to achieve network-level goals: the distributed Nash equilibrium (NE) seeking\cite{salehisadaghiani2016distributed,parise2015network,ye2017distributed}, best-response dynamics\cite{hopkins1999note,matsui1992best,nisan2011best}, and no-regret learning\cite{lin2021optimal,Bravo2018Bandit,zhou2018learning,gordon2008no}. 
	
	What these frameworks have in common is that players need to share information with others in a dynamic process, such as their actions, payoff gradients, or payoffs, and then choose their actions for the next stage based on the information received and their own payoff functions. Clearly, players' payoff functions are encoded in the shared information. However, players' payoff functions are often sensitive and private\cite{harsanyi1977morality}. As a result, players' payoff functions are at risk of privacy leakage. Owing to differential privacy\cite{dwork2008differential,dwork2014algorithmic}, it is possible for players to share information and decide their actions over time to achieve the desired outcome while keeping their payoff functions from being compromised. Differentially private systems have been well studied in the sense that lots of privacy algorithms are designed for various tasks such as average consensus\cite{huang2012differentially,nozari2017differentially}, estimation and filtering\cite{le2013differentially}, and convex optimization\cite{han2016differentially,huang2015differentially,nozari2018differentially}. As for differentially private games, the works of \cite{ye2021differentially,shakarami2022distributed} have focused on privacy-preserving distributed Nash seeking strategy design for aggregated games.
	\medskip
	
	{\noindent \bf Problem of Interest\quad} In this paper, we consider a network game where players are interconnected through an interaction/communication network. Players are involved in information aggregation processes that requires them to share information to accomplish certain collective goal. The shared information that encodes the sensitive information of payoff functions is monitored by adversaries. As a result, we aim to protect the differential privacy of players' payoff functions. 
	
	We are inspired by \cite{cortes2016differential,nozari2018differentially}. We propose a Laplace linear-quadratic functional perturbation (LLQFP) mechanism, which perturbs players' original payoff functions with linear-quadratic functional perturbation. The coefficients of those perturbation are generated by truncated Laplace distributions. The idea  is to let players participate in certain information aggregation process using the perturbed payoff functions. If the LLQFP mechanism preserves differential privacy, then it also enforces differential privacy of information aggregation processes by the resilience to post-processing of differential privacy\cite{dwork2014algorithmic}. 
	
	In the literature of differentially private information aggregation processes, a common approach is to add noises to players' shared information \cite{huang2015differentially,han2016differentially,zhang2018admm,ye2021differentially,shakarami2022distributed}. For this approach, perturbation has to be designed in accordance with a diverse set of objectives during information aggregation processes. Moreover, perturbation has to be added at all time steps, and therefore the longer the operating time of information aggregation processes is, the more amount of perturbation is required to add. Functional perturbation is easier to implement since its design does not depend on specific tasks. In addition, functional perturbation only adds perturbation once to produce the perturbed payoff functions, regardless of the number of steps players participate in the following information aggregation processes. 
	
	Functional perturbation was also proposed by \cite{cortes2016differential,nozari2018differentially}. They studied the distributed optimization problem subject to the requirement of differential privacy. Their work decomposed the objective functions into an infinite sequence of coefficients corresponding to the elements of a orthogonal basis in a separable Hilbert space, and added noises to the infinite coefficient sequence. Unfortunately, truncation is inevitable in practical implementations. Our work focuses on generalizing functional perturbation to the differentially private game setting. Instead of considering infinite expansion, we propose a mechanism that does not involve the decomposition of function space, but directly apply linear-quadratic functions  as perturbation avoiding the truncation problem.
	\medskip
	
	{\noindent \bf Contributions\quad}In this paper, we study network games under the differential privacy requirement for players’ payoff functions. We make the following contributions:
	\begin{itemize}
		\item We extend the notion of differential privacy to the network game setting, and propose a Laplace linear-quadratic functional perturbation (LLQFP) algorithm, which perturbs players' original payoff functions with linear-quadratic functional perturbation whose coefficients are generated according to truncated Laplace distributions.
		\item For monotone games, we show that the LLQFP algorithm maintains the concavity property of the perturbed payoff functions and yields a $\gamma$-accurate perturbed NE whose distance from the original NE is upper bounded by any prescribed constant $\gamma$.
		\item We investigate LQ games that players' payoff functions are parameterized. It serves as a tutorial example showing how Laplace parameters are selected to ensure certain differential privacy requirement.
		\item Experiments are conducted to verify the advantages of the LLQFP algorithm. 
	\end{itemize}
	\medskip
	
	{\noindent \bf Organization\quad}The remainder of the paper is organized as follows. For privacy concerns about players' payoff functions in network games, we formalize our problem in Section~\ref{sec:problem-formulation}.
	In Section~\ref{sec:propsed-algorithm}, we propose the LLQFP algorithm. In Section~\ref{sec:monotone-games}, we consider monotone games, and show the advantages of the LLQFP algorithm. In Section~\ref{sec:LQ Games}, we consider LQ games, and investigate Laplace parameter conditions that can guarantee certain different privacy requirement. Numerical examples are presented in Section~\ref{sec:numerical-examples}. This paper ends with concluding remarks in Section~\ref{sec:conclusion}.  
	
	\section{Problem Formulation}\label{sec:problem-formulation}
	\subsection{Network Games}
	Consider a network game with $n$ players. The players are interconnected through an interaction/communication network. The interaction/communication network is associated with a graph $\rmG(\rmV,\rmE)$, where $\rmV:=\{1,2,\dots,n\}$ represents the nodes (players), and $\rmE$ defines the links (the interdependency among players). Each player $i \in \rmV$ holds an action $x_{i} $ from a compact convex action space $\mcA_{i} \subseteq \mbR$. The aggregated action profile of all players and the action profile excluding player $i$ are denoted by $\bfx := [x_{1}, \dots, x_{n}]^{\top}$ and $\bfx_{-i} = [x_{1}, \dots,x_{i-1},x_{i+1},\dots, x_{n}]^{\top}$, respectively. Each player $i$ then receives her payoffs determined by a payoff function, i.e., $u_{i} = f_{i}(x_{i},\bfx_{-i}),$
	where the payoff function $f_{i} \in C^{2}(\mcA)$ is twice continuously differentiable over $\mcA:= \Pi_{i \in \rmV}\mcA_{i}$. 

	
	A common solution concept in game theory is called Nash equilibrium. It depicts an action profile under which no player may gain by simply modifying her action while others maintain theirs unaltered. We denote the NE by $\bfx^{\ast}:=[x_{1}^{\ast}, \dots, x_{n}^{\ast}]^{\top}$.
	
	\begin{definition}[Nash equilibrium] \label{def:ne}
		An action profile $\bfx^{\ast}$ is said to be a pure-strategy NE of a game if 
		$f_{i}(x_{i}^{\ast},\bfx_{-i}^{\ast}) \geq f_{i}(x_{i},\bfx_{-i}^{\ast}),\forall x_{i} \in \mcA_{i}, \forall i \in \rmV.$
	\end{definition}
	\medskip
	
	{\noindent \bf Information Aggregation Processes\quad}In network games, there are many network-level information aggregation operations that require players to share dynamical states over a horizon $t \in \{0,1,\dots,T\}$ to accomplish collective goals such as the distributed Nash seeking\cite{salehisadaghiani2016distributed,parise2015network,ye2017distributed}, best-response dynamics\cite{hopkins1999note,matsui1992best,nisan2011best}, and no-regret learning\cite{lin2021optimal,Bravo2018Bandit,zhou2018learning,gordon2008no}. 
	\medskip
	
	{\noindent \bf Example 1} (Distributed Nash seeking\cite{salehisadaghiani2016distributed,parise2015network,ye2017distributed})
	At time $t$, each player $i$ holds a dynamical state $\bfy_{i}(t)$ that typically consists of her action and her estimate of other players' actions. Then, each player $i$ shares $\bfy_{i}(t)$ with other players via certain interaction/communication network. Next, each player $i$ updates her dynamical state for $\bfy_{i}(t+1)$ based on the received players' dynamical states and her own payoff function $f_{i}$. The network-level objective is for (perhaps part of) the sequence $[\bfy_{1}(t); \dots;\bfy_{n}(t)], t = 0,1,\dots$  to converge to a NE.
	\medskip
	
	{\noindent \bf Example 2} (Best-response Dynamics\cite{hopkins1999note,matsui1992best,nisan2011best})
	At time $t$, each player $i$ holds a dynamical state $\bfy_{i}(t)$ that represents her action. Then, each $\bfy_{i}(t)$ is observed by or communicated with other players. Next, each player $i$ updates her state $\bfy_{i}(t+1)$ as the action that maximizes her payoff function given other players' current actions. Best-response dynamics is a behavioral model depicting how players strategically make decisions in a sequential manner. Sometimes, best-response dynamics converge to a NE. 
	\medskip
	
	{\noindent \bf Example 3} (No-regret Learning\cite{lin2021optimal,Bravo2018Bandit,zhou2018learning,gordon2008no})
	At time $t$, each player $i$ holds a dynamical state $\bfy_{i}(t)$ that typically consists of her action and her estimate of other players' payoff gradients. Then, each player $i$ shares $\bfy_{i}(t)$ with other players via certain interaction/communication network. Next, each player $i$ updates her dynamical state upon the received players' dynamical states and her own payoff function $f_{i}$. The network-level objective of no-regret learning is for the sequence $\bfy_{i}(t), t = 0,1,\dots$ to minimize the regret of player $i$ as the cumulative loss compared with a plain/single action in hindsight.

	\subsection{Problem Definition}
	{\noindent \bf Differentially Private Information Aggregation Processes\quad} From the above network-level information aggregation operations, it is clear that the $\bfy_{i}(t), i = 1,\dots, n, t = 0,1,\dots,T$ encode the information of payoff functions. Those states $\bfy_{i}(t)$ are shared by player $i$ with other players. However, players' payoff functions are often private and contains sensitive information \cite{harsanyi1977morality}. As a result, payoff functions face privacy risk in the information aggregation processes. 
	
	Differential privacy has been a standard tool to protect an individual's data privacy in a system where aggregate information is publicly published, but individual information is privately withheld \cite{dwork2014algorithmic}. Specifically, to protect the differential privacy of players' payoff functions, the mapping from $\bff := [f_{1};\dots;f_{n}]$ to  $\mathbf{Y} := [\bfy_{1}(0);\dots;\bfy_{1}(T);\dots;\bfy_{n}(0);\dots;\bfy_{n}(T)]$ should satisfy the following differential privacy condition. 
    \begin{definition}[$\mathcal{W}$-adjacency \cite{cortes2016differential}]\label{mu-adjacency}
    Given any normed vector space $(\mathcal{W}, || \cdot ||_{\mathcal{W}})$,
	$\bff$ and $\bff^{'}$ are said be $\mathcal{W}$-adjacent if there exists $i_{0} \in \rmV$ such that 
	\begin{subequations}
		\begin{align}
			&f_{i}=f^{'}_{i}\,,\qquad  i \neq i_{0}; \label{eq:differ-one}\\  
			&f_{i_{0}}-f^{'}_{i_{0}} \in \mathcal{W}. \label{eq:mu-adj}
		\end{align}
	\end{subequations}
	\end{definition}
	\noindent The normed vector space $\mathcal{W}$ is a design choice that we specify later according to the class of payoff functions. 
	\begin{definition}[$(\epsilon,\delta)$-differential privacy] \label{def:dp}
	The mapping $\mathcal{M}$ is said to preserve $(\epsilon,\delta)$-differential privacy if for any subset $\mathscr{M} \subseteq \range\big(\mathcal{M}\big)$,
	\begin{equation} \label{eq:dp}
		\mathbb{P}(\mathcal{M}(\bff) \in \mathscr{M}) \leq e^{\epsilon} \mathbb{P}(\mathcal{M}(\bff') \in \mathscr{M}) + \delta,      
	\end{equation}
	holds for any two $\mathcal{W}$-adjacent payoff functions $\bff$ and $\bff'$. 
	\end{definition}

	\subsection{Functional Perturbation}	
	We propose a functional perturbation mechanism from $\bff$ to $\hat{\bff}$ where certain perturbation is added to produce $\hat{\bff}$, and then players use $\hat{\bff}$ to participate in information aggregation processes. As a result, the privacy of players' payoff functions $\bff$ may be protected in the sense that Definition~\ref{def:dp} may be satisfied. If the mapping from $\bff$ to $\hat{\bff}$ preserves differential privacy, then differential privacy of the mapping from $\bff$ to $\mathbf{Y}$ is also enforced by the immune to post-processing \cite{dwork2014algorithmic}.
	
	There are a few practical challenges in designing such a mechanism:
	\begin{itemize}
		\item Differential privacy of the functional perturbation mechanism from $\bff$ to $\hat{\bff}$ should be provable.
		\item The basic regularity property of the game should be maintained. In particular, if $\bff$ are concave, $\hat{\bff}$ should be also concave. 
		\item The distance between the NE of the original game and the NE of the perturbed game should be upper bounded and adjustable by parameter tuning.
	\end{itemize}
	
	In this paper, we aim to develop a distributed algorithm to realize this functional perturbation mechanism that can address the above challenges.

	\section{The Proposed Algorithm}\label{sec:propsed-algorithm}
	 The truncated Laplace distribution truncated by $[-a,a]$ with mean zero and scale parameter $\lambda$, denoted $\mathscr{L}_{tr}(a,\lambda)$, has probability density function 
	\begin{equation}\label{eq:truncated-Laplace}
		p(x;a,\lambda)=\left\{
		\begin{array}{ll} Be^{-|x|/\lambda},\quad &\mbox{for } x\in[-a,a], \\ 
			0\,,\quad & \mbox{otherwise,} 
		\end{array} \right.
	\end{equation}
	where $B = \frac{1}{2\lambda(1-e^{-a/\lambda})}.$
	
	Denote the neighbors of player $i$ by the set $\rmN_{i} \subset \rmV$. We sort the indices of player $i$'s neighbors  in ascending order in the set $\rmO_{i} :=\{i_{1},i_{2},\dots,i_{|\rmN_{i}|}\}$. For example, if player $j$ is player $i$'s $k$th neighbor, then $i_{k} = j.$ 
	
	\subsection{LLQFP Algorithm}
	We next propose a Laplace linear-quadratic functional perturbation Algorithm in Algorithm~\ref{alg:laplace_lq}.
	
	\begin{algorithm}[h]
		\caption{Laplace Linear-quadratic Functional Perturbation Algorithm}
		\begin{algorithmic}[1]
			\label{alg:laplace_lq}
			\renewcommand{\algorithmicensure}{\textbf{Input:}}	
			\ENSURE   Laplace parameters $a,\lambda$; payoff functions $f_{1},\dots,f_{n};$
			\renewcommand{\algorithmicensure}{\textbf{Output:}}	
			\ENSURE  perturbed payoff functions $\hat{f_{1}}, \dots, \hat{f_{n}}$
			
			\STATE Each player $i \in \rmV$ independently generates a  sequence random numbers $\omega_{i,k}$, $k=1,\ldots,|\rmN_i|+2$, according to $ \mathscr{L}_{tr}(a,\lambda)$ in \eqref{eq:truncated-Laplace}.
			\STATE Each player $i \in \rmV$ computes
			\begin{subequations}
				\begin{align}
					& q_{ij} =\left\{
					\begin{array}{ll} \frac{\omega_{i,(|\rmN_{i}|+1)}}{2}+\frac{a(|\rmN_{i}|+1)}{2}, \quad & j = i, \\ 
						\omega_{i,k}, \quad  &j = i_{k},\\
						0,\quad & \mbox{otherwise,} 
					\end{array} \right. \label{eq:q_ij}\\
					&\beta_{i}=\omega_{i,(|\rmN_{i}|+2)}. \label{eq:beta-i}
				\end{align}
			\end{subequations} 
			\STATE Each player $i \in \rmV$ employs a perturbed payoff function based on $\bfq_{i}:=[q_{i1},\dots,q_{in}]^{\top}$ and $\beta_{i}$:
			\begin{equation}\label{eq:perturbed-payoff}
				\hat f_{i}(x_{i},\bfx_{-i})= f_{i}(x_{i},\bfx_{-i}) - x_{i} \bfq_{i}^\top\bfx - \beta_{i} x_{i}.
			\end{equation}
			\RETURN $\hat{f_{1}}, \dots, \hat{f_{n}}$
		\end{algorithmic}
	\end{algorithm}
	
	Algorithm~\ref{alg:laplace_lq} perturbs the original payoff functions with linear-quadratic functional perturbation whose coefficients are generated according to \eqref{eq:truncated-Laplace}. Each player $i$ first generates $|\rmN_{i}|+2$ independent truncated Laplace noises, and then strategically inserts noises to her payoff function in a linear-quadratic perturbation form. Each $\bfq_{i}$ in the perturbed payoff function is not a simple stack of $|\rmN_{i}|+1$ truncated Laplace noises, but depends on the network structure and the truncated bound $a$. 
	
	 The work of \cite{cortes2016differential,nozari2018differentially} also presented an analysis on the the mechanism of functional perturbation via Laplace noises under the problem setting of differentially private distributed convex optimization. The differences between theirs and ours are as follows:
	\begin{itemize}
		\item Their work was based on the assumption that the objective functions are twice continuously differentiable functions with bounded gradients and Hessians. But we are going to assume that the underlying game is a monotone game, which is common in the game literature \cite{rosen1965existence}. As a result, investigating whether the basic properties of the game is maintained and characterizing the accuracy of NE after applying Algorithm~\ref{alg:laplace_lq} are different form their framework.
		
		\item Their work decomposed $\bff$ into its coefficients by the infinite expansion, and perturbed this infinite sequence by adding noise to all of its elements. As a result, truncation is inevitable in practical implementations. But we only consider linear-quadratic functional  perturbation where the noise coefficients are related with the underlying interaction graph of the game. In a sense, the network structure of the game may be maintained. 
		
	\end{itemize}
	
	\subsection{Positivity Guarantee}
	We now present a property of the coefficients generated according to $\mathscr{L}_{tr}(a,\lambda)$, which is necessary for the Theorems later.
	
	Denote $\bfd_{i} = [q_{i1},\dots,q_{i(i-1)},2q_{ii},q_{i(i+1)},\dots,q_{in}]^{\top}$ and $\bfD = [\bfd_{1} \ \bfd_{2} \ \dots \ \bfd_{n}]$. Also denote $\betab = [\beta_{1},\dots,\beta_{n}]^{\top}$.

	\begin{lemma} \label{lemma:Q-PSD}
		$\bfD^{\top}$ is a positive semidefinite matrix.
	\end{lemma}
	\noindent{\it Proof.} We focus on the magnitude of the diagonal element in each row, and the sum of the magnitudes of all non-diagonal elements in that row. According to~\eqref{eq:q_ij}, we have $|2q_{ii}| \in [a|\rmN_{i}|,a(|\rmN_{i}|+2)]$ and $\sum_{i \neq j}|q_{ij}| \in [0,a|\rmN_{i}|], \forall i \in \rmV$. Since $|2q_{ii}| \geq \sum_{i \neq j}|q_{ij}|$, $ \bfD^{\top}$ is diagonally dominant. A symmetric diagonally dominant real matrix with nonnegative diagonal entries is positive semidefinite. Hence, $ \bfD^{\top}$ is a positive semidefinite matrix. \hfill$\square$
	\medskip

	\section{Monotone Games}\label{sec:monotone-games}
	In what follows, we look at a class of strongly monotone games, present its basic properties, and  show the advantages of Algorithm~\ref{alg:laplace_lq}.
	
	For each $i \in \rmV$, we denote the gradient of $f_{i}$ with respect to $x_{i}$ by $\nabla_{x_{i}} f_{i} :=\frac{\partial f_{i}}{\partial x_{i}} \in \mathbb{R}$, and $\phib(\bfx):=[\nabla_{x_{1}}f_{1},\dots,\nabla_{x_{n}}f_{n}]^{\top} \in \mathbb{R}^{n}$. 
	
	We impose the following assumption of a class of strongly monotone games.
	\begin{assumption}[\hspace{1sp}\cite{ye2021differentially}]\label{apt:strongly-monotone-games}
		For some $l_{m} >0$ and for all $\bfx',\bfx \in \mcA,$
		\begin{equation}\label{eq:strongly-montone-games}			
			\sum_{i \in \rmV} (\phi_{i}(\bfx)-\phi_{i}(\bfx'))(x_{i}-x'_{i}) \leq -l_{m}\|\bfx - \bfx'\|^{2}. 
		\end{equation}
	\end{assumption}
	Assumption~\ref{apt:strongly-monotone-games} implies that each player's original payoff function $\bff$ is strictly concave in $x_{i}$\cite{rosen1965existence}. We introduce the definition of concavity preservation.
	\begin{definition}[Concavity preservation \cite{cortes2016differential}] \label{def:cp}
		Let Assumption~\ref{apt:strongly-monotone-games} hold. Algorithm~\ref{alg:laplace_lq} is said to be concavity-preserving if each $\hat{f}_{i}$ is strictly concave  in $x_{i}$ for all $i \in \rmV$. 
	\end{definition}
	
	\subsection{Concavity Preservation}
	The following result proves that under Assumption~\ref{apt:strongly-monotone-games}, Algorithm~\ref{alg:laplace_lq} is concavity-preserving. 

	\begin{theorem}\label{thm:concavity-preservation}
		Let Assumption~\ref{apt:strongly-monotone-games} hold. Then, Algorithm~\ref{alg:laplace_lq} is concavity-preserving. 
	\end{theorem}
	\noindent {\it Proof.} Under Assumption~\ref{apt:strongly-monotone-games}, each player $i$'s original payoff function is strictly concave in $x_{i}$. We consider $\bfx = [x_{1},\dots,x_{i},\dots,x_{n}]^{\top}$ and $\bfx' = [x_{1},\dots,x'_{i},\dots,x_{n}]^{\top}$ and obtain $(\phi_{i}(\bfx)-\phi_{i}(\bfx'))(x_{i}-x'_{i})<0.$ 
	
	We then check the sign of $(\hat{\phi}_{i}(\bfx)-\hat{\phi}_{i}(\bfx'))(x_{i}-x'_{i})$ = $(\phi_{i}(\bfx)-\phi_{i}(\bfx'))(x_{i}-x'_{i})-2q_{ii}(x_{i}-x'_{i})^{2}<0$,
	which complies with Definition~\ref{def:cp}. \hfill$\square$
	\medskip
	
	The next result shows that under Assumption~\ref{apt:strongly-monotone-games}, both original game and perturbed game after Algorithm~\ref{alg:laplace_lq} admit a unique NE.
	\begin{theorem}\label{thm:unique-both-ne}
	Let Assumption~\ref{apt:strongly-monotone-games} hold. Then,
	
	(i) the original game with the payoff functions $\bff$ admits a unique NE.
	
	(ii) after Algorithm~\ref{alg:laplace_lq}, the perturbed game with the perturbed payoff functions $\hat{\bff}$ admits a unique NE. 
	\end{theorem}
	\noindent{\it Proof.} For (i), the class of strongly monotone games is a proper subclass of monotone games, first introduced in \cite{rosen1965existence}. Instead of the stronger requirement in Assumption~\ref{apt:strongly-monotone-games}, the weaker assumption $	\sum_{i \in \rmV} c_{i} (\phi_{i}(\bfx) - \phi_{i}(\bfx'))(x_{i} - x'_{i}) < 0$ is imposed. Every monotone game admits a unique NE \cite[Theorem 2]{rosen1965existence}. Therefore, the original game under Assumption~\ref{apt:strongly-monotone-games} also admits a unique NE.  
	
	For (ii), we are going to check whether the perturbed game is a monotone game by investigating whether the sign of $\sum_{i \in \rmV} (\hat{\phi_{i}}(\bfx)-\hat{\phi_{i}}(\bfx'))(x_{i} - x'_{i}) < 0,\forall \bfx',\bfx \in \mcA, \bfx'\neq\bfx.$ It is straightforward that
		\begin{align*}
			& \sum_{i \in \rmV} (\hat{\phi_{i}}(\bfx)-\hat{\phi_{i}}(\bfx'))(x_{i} - x'_{i})\\
			=& \sum_{i \in \rmV} (\phi_{i}(\bfx)-\phi_{i}(\bfx'))(x_{i} - x'_{i}) - \sum_{i \in \rmV}2q_{ii}(x_{i} - x'_{i})^{2} < 0.
		\end{align*}
	We draw the conclusion that the perturbed game is a monotone game, therefore admitting a unique NE. \hfill$\square$
	
	\subsection{$\gamma$-accurate Nash equilibrium}
	According to Theorem~\ref{thm:unique-both-ne}, when Assumption~\ref{apt:strongly-monotone-games} holds, both original game and perturbed game admit a unique NE. We denote the original NE of the original game by $\bfx^{\ast}$ and the perturbed NE of the perturbed game by $\hat{\bfx}^{\ast}$.
	
	We now introduce the definition of $\gamma$-accurate NE. 
	\begin{definition}[$\gamma$-accurate NE] \label{def:gamma-pne}
	Let Assumption~\ref{apt:strongly-monotone-games} hold. The perturbed NE $\hat{\bfx}^{\ast}$ is said to be $\gamma$-accurate if $\|\bfx^*-\hat{\bfx}^*\|\leq\gamma$. 
	\end{definition}
	\medskip
	
	In the following result, we derive an upper bound for the distance between the original NE and the perturbed NE after Algorithm~\ref{alg:laplace_lq}.
	\begin{theorem}\label{thm:gamma_bound}
	Let Assumption~\ref{apt:strongly-monotone-games} hold. Further suppose that the original NE and perturbed NE are both interior points in the action space $\mcA$. Then, the perturbed NE is $\gamma$-accurate with 
	\begin{equation}\label{eq:gamma_bound1}
		\gamma = \frac{\sqrt{n}a+\sqrt{\sum_{i\in \rmV}(4|\rmN|^{2}_{i}+5|\rmN_{i}|+4)} a\|\bfx^{\ast}\|}{l_{m}}
	\end{equation}
	\end{theorem}
	\noindent {\it Proof.} By taking the derivative of \eqref{eq:perturbed-payoff} w.r.t. $x_{i}$ and rewriting it in the vector form, we obtain 
	\begin{equation}\label{eq:vector-perturbed-derivatives}
		\hat{\phib}(\bfx) = \phib(\bfx) - \bfD^{\top}\bfx - \betab.   
	\end{equation}
	We now turn to the original NE and perturbed NE, whose existence and uniqueness are guaranteed by Theorem~\ref{thm:unique-both-ne}. Moreover, in Theorem~\ref{thm:gamma_bound}, we further impose the assumption of interior NE. Substituting $\hat{\bfx}^{\ast}$ into Eq.~\eqref{eq:vector-perturbed-derivatives}, we have
	$\phib(\hat{\bfx}^{\ast}) = \hat{\phib}(\hat{\bfx}^{\ast}) + \bfD^{\top}\hat{\bfx}^{\ast} + \betab.$
	The first-order condition for the interior NE is that 
	$\phib(\bfx^{\ast}) = 0$ and $\hat{\phib}(\hat{\bfx}^{\ast}) = 0.$ 
	Then, it yields $
	\big[\phib(\bfx^{\ast}) - \phib(\hat{\bfx}^{\ast})\big] - \bfD^{\top}(\bfx^{\ast} - \hat{\bfx}^{\ast} ) = -\betab - \bfD^{\top}\bfx^{\ast} .$       
	Multiplying by $(\bfx^{\ast} - \hat{\bfx}^{\ast})^{\top}$, we observe that 
	\begin{align*}
		&-(\bfx^{\ast} - \hat{\bfx}^{\ast})^{\top}(\betab + \bfD^{\top}\bfx^{\ast})\\
		=& (\bfx^{\ast} - \hat{\bfx}^{\ast})^{\top}\big[\phib(\bfx^{\ast}) - \phib(\hat{\bfx}^{\ast})\big] - (\bfx^{\ast} - \hat{\bfx}^{\ast})^{\top}\bfD^{\top}(\bfx^{\ast} - \hat{\bfx}^{\ast} )\\
		\overset{(a)}{\leq}&  (\bfx^{\ast} - \hat{\bfx}^{\ast})^{\top}\big[\phib(\bfx^{\ast}) - \phib(\hat{\bfx}^{\ast})\big] \\
		\overset{(b)}{\leq} &- l_{m}\|\bfx^{\ast} - \hat{\bfx}^{\ast}\|^{2} . 
	\end{align*}        
	The inequality (a) holds because $\bfD^{\top}$ is designed to be a positive semidefinite matrix (See Lemma~\ref{lemma:Q-PSD}). The inequality (b) is exactly from Eq.~\eqref{eq:strongly-montone-games}. Thus there holds $
	(\bfx^{\ast} - \hat{\bfx}^{\ast})^{\top}(\betab + \bfD^{\top}\bfx^{\ast}) \geq  l_{m}\|\bfx^{\ast} - \hat{\bfx}^{\ast}\|^{2} > 0 .$
	
	Further considering the Cauchy–Schwarz inequality, we finally obtain a bound on the distance  
	\begin{subequations}
		\begin{align}
			&\| \bfx^{\ast}-\hat{\bfx}^{\ast}\|\\ 
			\leq & \frac{\|\betab\|+\|\bfD\|\|\bfx^{\ast}\|}{l_{m}} \label{eq:gamma2} \\
			\leq & 
			\frac{\sqrt{n}a+\sqrt{\sum_{i\in \rmV}(4|\rmN|^{2}_{i}+5|\rmN_{i}|+4)} a\|\bfx^{\ast}\|}{l_{m}}. \label{eq:gamma1}
		\end{align}
	\end{subequations}
	The proof is now completed.$\hfill \square$
	\begin{remark}\label{rmk:gamma}
		The upper bound~\eqref{eq:gamma1} is very conservative because it considers the worst-case scenario in which all $q_{ij}, \beta_{i}, \forall i, j \in \rmV$ take their maximum values. In comparison, the upper bound~\eqref{eq:gamma2} is a relatively small bound, which is more likely to happen with high probability in realization.
	\end{remark}
	\section{Linear-quadratic Games}\label{sec:LQ Games}
	A practical challenge of Algorithm~\ref{alg:laplace_lq} is how to select the Laplace parameters $a$ and $\lambda$ that can ensure certain differential privacy requirement. The selection depends on the class of payoff functions and the design choice of $\mathcal{W}$-adjacency. In this section, we analyze a benchmark game whose payoff functions are in the linear-quadratic form.
	
	Denote the adjacency matrix of the interaction/communication network by $\bfG \in \mathbb{R}^{n\times n}$, with each entry $g_{ij} \in \mbR$ denoting whether player $j \in \rmV$ is linked to player $i \in \rmV$ and also indicating the linkage intensity.

	We now impose the assumption of linear-quadratic games.
	\begin{assumption}\label{apt:lq}
	The payoff functions of a linear-quadratic game are set as
	\begin{equation}\label{eq:linear-quadratic payoffs}
	f_{i}(x_{i},\bfx_{-i}) = -\frac{1}{2}x_i^2+b_ix_i+\sum_{j\in \rmV}
	g_{ij}x_ix_j, \quad \forall i \in \rmV,
	\end{equation}
	where $b_{i} \in \mbR^{\geq 0}$ represents the marginal benefit of player $i$.
	\end{assumption}	
	
	Since each player $i$'s payoff functions is now parameterized by the parameters $g_{ij}$ and $b_{i}$. Hence, it is reasonable to specify $\mathcal{W}$-adjacency on these parameters. For example, we specify a definition of $\mu$-adjacency for LQ games.
	\begin{definition}[$\mu$-adjacency]\label{def:mu-adjacency}
	Consider linear-quadratic payoff functions $\bff$ and $\bff'$. They are said to be $\mu$-adjacent if there exists $i_{0} \in \rmV$ such that
	\begin{subequations}
	\begin{align}
	&g_{i1}=g'_{i1}, \dots, g_{in}=g'_{in},  b_{i} = b'_{i}, \quad i \neq i_{0}; \label{eq:differ-one}\\  
	&\max\{g_{i_{0}1}-g'_{i_{0}1},\dots,g_{i_{0}n}-g'_{i_{0}n},b_{i_{0}}-b'_{i_{0}}\}\leq \mu.\label{eq:mu-adj}
	\end{align}
	\end{subequations}
	\end{definition}		
	\medskip
	
	In what follows, we first present how the Laplace parameters $a$ and $\lambda$ are selected to guarantee $(\epsilon,\delta)$-differential privacy for one-dimensional truncated Laplace mechanism, and then generalize this result to differential private LQ games.
	
	\subsection{One-dimensional Truncated Laplace Mechanism}
	
	The work of \cite{geng2020tight,croft2019differential,holohan2018bounded} investigated how the Laplace parameters are chosen to meet the differential privacy criterion in the one-dimensional case. As presented in \eqref{eq:gamma1}, Algorithm~\ref{alg:laplace_lq} leads to a biased perturbed NE. As a result, a stringent analysis is required to determine the lower bounds of the Laplace parameters that can produce a less biased perturbed NE. Compared with \cite{geng2020tight}, the following result relaxes the requirement for the Laplace parameters to guarantee the differential privacy, and gives the tight lower bounds for $a$ and $\lambda$ in the one-dimensional case.
	
	We consider one-dimensional truncated perturbation mechanism.
	Let $D$ be the space of datasets of interest. Suppose there is a query as a function $y:D \to \mathbb{R}$.  Given $D$, a randomized mapping $\mathcal{K}$ will release the one-dimensional response $\mathcal{K}(D)$ that is the summation of the true query answer $y \in \mathbb{R}$ and a random noise $\eta \in [-a,a]$ following $\mathscr{L}_{tr}(a,\lambda)$, $\mathcal{K}(D) = y(D) + \eta$. 	The sensitivity of one-dimensional true query is then given by $\Delta y = \max_{D_{1},D_{2} \in D} |y(D_{1}) - y(D_{2})|.$ The randomized mapping $\mathcal{K}$ gives $(\epsilon,\delta)$-differential privacy if for any two datasets $D_{1},D_{2} \in D$ differing in at most one element, and all $\mathscr{K} \subseteq \range(\mathcal{K})$, there holds
	\begin{equation}\label{eq:dwork-dp}
		\mathbb{P}(\mathcal{K}(D_{1})\in \mathscr{K}) \leq e^{\epsilon} \mathbb{P}(\mathcal{K}(D_{2})\in \mathscr{K}) + \delta.
	\end{equation}

	\begin{lemma}\label{lemma:1-d-dp}
		Given the privacy parameters $0<\delta<\frac{1}{2},\epsilon >0$, the randomized mapping $\mathcal{K}$ preserves $(\epsilon,\delta)$-differential privacy  if
		\begin{subequations}
			\begin{align}
				&\lambda \geq \frac{\Delta y}{\epsilon - \ln(1-\delta)}, \label{eq:lambda-condition}\\
				&a \geq  \max \Big\{\Delta y, \lambda \ln\left(\frac{e^{\frac{\Delta y}{\lambda}}-1}{2\delta}+1\right)\Big\}. \label{eq:a-condition}
			\end{align}
		\end{subequations}
	\end{lemma}
	\noindent{\it Proof.} We are seeking to show that for any $D_{1},D_{2} \in D$ differing in at most one element, for any subset $\mathscr{K} \subseteq \range(\mathcal{K}),$ Eq.~\eqref{eq:dwork-dp} is satisfied. 
	
	Without loss of generality, we let $y(D_{1}) \leq y(D_{2})$. Given $\mathscr{K} \subseteq \range(\mathcal{K})$, there are $5$ cases to consider, each of which should render Eq.~\eqref{eq:dwork-dp} to be satisfied.
	\begin{itemize}
		\item[1.] $\mathscr{K} \subseteq (-\infty, -a + y(D_{1})]:$ It is true that $0\leq e^{\epsilon}\cdot0 + \delta$.
		
		\item[2.] $\mathscr{K} \subseteq [-a + y(D_{1}), -a + y(D_{2})]:$ First, since $\delta < \frac{1}{2}$, it is impossible to find the configuration for $a$ and $\lambda$ that can make Eq.~\eqref{eq:dwork-dp} valid when $a \leq \Delta y$. Second, we now consider $y(D_{2})-y(D_{1}) \leq \Delta y < a$. For any $D_{1}$ and $D_{2}$ differing in at most one element, to satisfy Eq.~\eqref{eq:dwork-dp}, we are going to show that the probability mass in the interval $[-a+y(D_{1}),-a+y(D_{2})]$ does not exceed $\delta$:
		\begin{align*}
			\int_{\mathscr{K}}Be^{\frac{-|y-y(D_{1})|}{\lambda}}dy = & B\lambda(e^{\frac{-a+y(D_{2})-y(D_{1})}{\lambda}}-e^{\frac{-a}{\lambda}})\\
			\leq & B\lambda(e^{\frac{-a+\Delta}{\lambda}}-e^{\frac{-a}{\lambda}}) \leq  \delta.
		\end{align*}
		The first inequality holds because $e^{\frac{-a+y(D_{2})-y(D_{1})}{\lambda}}$ increases when $y(D_{2})-y(D_{1})$ increases, while the second inequality comes from the condition~\eqref{eq:a-condition}.
		
		\item[3.] $\mathscr{K} \subseteq [-a + y(D_{2}), a + y(D_{1})]:$ Equation~\eqref{eq:dwork-dp} can be written as
		$\int_{\mathscr{K}}Be^{\frac{-|y-y(D_{1})|}{\lambda}}dy \leq e^{\epsilon}\int_{\mathscr{K}}Be^{\frac{-|y-y(D_{2})|}{\lambda}}dy + \delta.$ Using triangle inequality  $|y-y(D_{2})|\leq|y-y(D_{1})|+|y(D_{1})-y(D_{2})|$ and combining the condition~\eqref{eq:lambda-condition}, it is sufficient to show that
		$\int_{\mathscr{K}}Be^{\frac{-|y-y(D_{1})|}{\lambda}}dy \leq e^{\epsilon-\frac{\Delta}{\lambda}}\int_{\mathscr{K}}Be^{\frac{-|y-y(D_{1})|}{\lambda}}dy + \delta,$ or further, $1 \leq e^{\epsilon-\frac{\Delta y}{\lambda}} + \delta \leq e^{\epsilon-\frac{\Delta y}{\lambda}} + \frac{\delta}{\int_{\mathscr{K}}Be^{\frac{-|y-y(D_{1})|}{\lambda}}dy}.$ 
		\item[4.] $\mathscr{K} \subseteq [a + y(D_{1}), a + y(D_{2})]:$ It is true that $0 \leq e^{\epsilon}\int_{\mathscr{K}}Be^{\frac{-|y-y(D_{2})|}{\lambda}}dy + \delta.$
		
		\item[5.] $\mathscr{K} \subseteq [a + y(D_{2}), \infty]:$ It is valid that $0\leq e^{\epsilon}\cdot0 + \delta$.
	\end{itemize}

	In all, under conditions~\eqref{eq:lambda-condition} and~\eqref{eq:a-condition}, the randomized mapping $\mathcal{K}$ preserves $(\epsilon,\delta)$-differential privacy. \hfill$\square$
	
	\subsection{Differentially Private LQ Games}
	In what follows, Laplace parameter conditions are given to ensure certain differential privacy requirement for LQ games.
	
	Stack the non-zero elements $q_{ij}$ into  $\bfq \in \mbR^{m}$ with $m = n + \sum_{i \in \rmV}|N_{i}|$. Also stack $g_{ij}$ into $\bfg \in \mbR^{m}$ such that each element in $\bfg$ is matched with the corresponding element in $\bfq$. In particular, if the $k$th element of $\bfq$ is $q_{ij}$, then the $k$th element of $\bfg$ is $g_{ij}$. The dimension of $\begin{bmatrix} \bfg \cr \bfb \end{bmatrix}$ is $l = 2n + \sum_{i \in \rmV}|N_{i}| $. Further define $p = 1 + \max_{i \in \rmV}|N_{i}|$. 
	
	\begin{theorem}\label{thm:a_lambda_threshold}
		Consider a LQ game. Then given any $\epsilon,\delta,\mu > 0$, the mapping $
		\widehat{\mathcal{M}}(\bfg,\bfb) = \begin{bmatrix} \bfg-\bfq \cr \bfb -\betab \end{bmatrix}$
		achieves $(p\epsilon,p\delta)$-differential privacy under $\mu$-adjacency if
		\begin{subequations}
			\begin{align}
				&\lambda \geq \frac{\mu}{\epsilon - \ln(1-\delta)}, \label{eq:thm-lambda-condition}\\
				&a \geq  \max \Big\{\mu, \lambda \ln\left(\frac{e^{\frac{\mu}{\lambda}}-1}{2\delta}+1\right)\Big\}. \label{eq:thm-a-condition}
			\end{align}
		\end{subequations}
	\end{theorem}
	\noindent{\it Proof.}
	Consider two $\mu$-adjacent linear-quadratic payoff functions $\bff$ and $\bff^{'}$ that are uniquely determined by the pairs $(\bfg,\bfb)$ and $(\bfg',\bfb')$, respectively.
	
	Denote $\bfv = [\bfg^{\top} \ \bfb^{\top}]^{\top} \in \mathbb{R}^{l}$ and $\bfv' = [\bfg'^{\top} \ \bfb'^{\top}]^{\top} \in \mathbb{R}^{l}.$ Also define $\rmV_{perturbed} = \{1,2,\dots,l\}$. Due to $\mu$-adjacency, there exists $i_{0} \in \rmV$ such that the conditions~\eqref{eq:differ-one} and \eqref{eq:mu-adj} hold. We denote by $\rmV_{diff}$ the indices of $b_{i_{0}}$ and $g_{i_{0}j},j\in \rmN_{i_{0}}\cup\{i_{0}\}$, in $\bfv$. The conditions~\eqref{eq:differ-one} and \eqref{eq:mu-adj}  indicate that 1) $\bfv$ and $\bfv'$ differ in at most $|\rmN_{i_{0}}|+1$ elements; 2) and for any $i \in \rmV_{diff},$ we have $ |v_{i} - v'_{i}| \leq \mu$. 
	
	Note that each $v_{i}, i \in \rmV_{diff},$ is independent of any other $v_{j}, j \in \rmV_{perturbed} \neq i$. We decompose $\widehat{\mathcal{M}}$ and further notice that each component $\widehat{\mathcal{M}}_{i}(v_{i}),i \in \rmV_{diff},$ can be viewed as a randomization of $v_{i}$. We then apply Lemma~\ref{lemma:1-d-dp} with $\Delta y = \mu$. It is therefore straightforward that when the conditions~\eqref{eq:thm-lambda-condition} and~\eqref{eq:thm-a-condition} are satisfied, each component $\widehat{\mathcal{M}}_{i}(v_{i}),i \in \rmV_{diff},$ preserves $(\epsilon,\delta)$-differential privacy. 
	
	We now examine the probability 
	$\mathbb{P}(\widehat{\mathcal{M}}(\bfv) \in \widehat{\mathscr{M}})=\prod_{i \in \rmV_{perturbed}} \mathbb{P}(\widehat{\mathcal{M}}_{i}(v_{i}) \in \widehat{\mathscr{M}}_{i}).$
	There are at most $|\rmN_{i_{0}}|+1$ different elements indexed in $\rmV_{diff}$ between $\bfv$ and $\bfv'$, while the remaining elements indexed in the set $\rmV_{same}:=(\rmV_{perturbed}-\rmV_{diff})$ are the same. Also, combining the fact that each component $\widehat{\mathcal{M}}_{i}(v_{i}),i \in \rmV_{diff},$ is $(\epsilon,\delta)$-differentially private, as a consequence, we can substitute
	\begin{align*}
		&\prod_{i \in \rmV_{perturbed}} \mathbb{P}(\widehat{\mathcal{M}}_{i}(v_{i}) \in \widehat{\mathscr{M}}_{i})\\
		=&\prod_{i \in \rmV_{same}} \mathbb{P}(\widehat{\mathcal{M}}_{i}(v_{i}) \in \widehat{\mathscr{M}}_{i}) \prod_{i \in \rmV_{diff}} \mathbb{P}(\widehat{\mathcal{M}}_{i}(v_{i}) \in \widehat{\mathscr{M}}_{i}) \\
		\leq& \prod_{i \in \rmV_{same}} \mathbb{P}(\widehat{\mathcal{M}}_{i}(v'_{i}) \in \widehat{\mathscr{M}}_{i}) \prod_{i \in \rmV_{diff}} (e^{\epsilon}\mathbb{P}(\widehat{\mathcal{M}}_{i}(v'_{i}) \in \widehat{\mathscr{M}}_{i}) + \delta) 
	\end{align*}
	If we focus on the second product term and look at the additive contribution of each of the $\delta$ terms, of which there are $|\rmN_{i_{0}}|+1$, we notice that they are only ever multiplied by probabilities that are at most one
	Therefore, each contributes at most an additive $\delta$:
	\begin{align*}
		&\prod_{i \in \rmV_{diff}} (e^{\epsilon}\mathbb{P}(\widehat{\mathcal{M}}_{i}(v'_{i}) \in \widehat{\mathscr{M}}_{i}) + \delta)\\
		\leq&  e^{(|\rmN_{i_{0}}|+1)\epsilon} \prod_{i \in \rmV_{diff}} \mathbb{P}(\widehat{\mathcal{M}}_{i}(v'_{i}) \in \widehat{\mathscr{M}}_{i}) + (|\rmN_{i_{0}}|+1)\delta 
	\end{align*}
	Then, we have
	\begin{align*}
		&\prod_{i \in \rmV_{perturbed}} \mathbb{P}(\widehat{\mathcal{M}}_{i}(v_{i}) \in \widehat{\mathscr{M}}_{i}) \\
		&\leq e^{(|\rmN_{i_{0}}|+1)\epsilon} \mathbb{P}(\widehat{\mathcal{M}}(\bfv') \in \widehat{\mathscr{M}}) + (|\rmN_{i_{0}}|+1)\delta.
	\end{align*}
	Note that $i_{0}$ can be any $i \in \rmV$. 
	Therefore, $\widehat{\mathcal{M}}$ releases $(p\epsilon,p\delta)$-differential privacy with $p = 1 + \max_{i \in \rmV}|N_{i}|$. \hfill$\square$
	\begin{remark}\label{rmk:lq_dp}
	Note that the privacy guarantee depends in a crucial way on how the notion of adjacency is defined. Although we only prove the differential privacy from $\bff$ to $\hat{\bff}$ for LQ games with $\mu$-adjacency in Definition~\ref{def:mu-adjacency}, it serves as a tutorial example and can be applied to other monotone games as long as players' payoff functions are explicitly given.
	\end{remark}
	\section{Numerical Examples}\label{sec:numerical-examples}
	Consider a LQ game with $10$ players. Players are arranged in a ring lattice with each player connected to $| \rmN_{i} |=4,\forall i \in \rmV$, neighbors. (Thus, $p = 5$.) Each linkage intensity is set as $0.08$. 
	The action space considered is relatively large, $\bfx_{i} \in [0,100],\forall i \in \rmV$, to ensure interior original NE and perturbed NE. 
	According to \cite{ballester2006s}, the original NE is calculated by $\bfx^{\ast} = (\mathbf{I}-\bfG)^{-1}\bfb$. 
	
	\begin{experiment}[Validation of Theorems]\label{exp:validations}
		\rm We consider  $\mu = 0.01$ and choose two parameter configurations 
		\begin{align}
			&\epsilon_{1} = \ln 2,  \   \delta_{1} = 0.05,  a_{1} = 0.034,   \ \lambda_{1} = 0.013;\tag{S1}\\
			&\epsilon_{2} = 3\ln 2,  \ \delta_{2} = 0.15,  a_{2} = 0.015,  \  \lambda_{2} = 0.0045,\tag{S2}
		\end{align} 
		where $a_{1},a_{2},\lambda_{1},\lambda_{2}$ are selected upon Theorem~\ref{thm:a_lambda_threshold} to guarantee differential privacy. 
		
		Under each parameter configuration, we conduct $500$ executions, in each of which we apply Algorithm~\ref{alg:laplace_lq}. Each perturbed NE is calculated by $\hat{\bfx}^{\ast} = (\mathbf{I}-\bfG+\bfD)^{-1}(\bfb - \betab)$. We then compute $\|\bfx^*-\hat{\bfx}^*\|$, and $\gamma$ according to \eqref{eq:gamma2}. 
		\begin{figure}[b]
		\centering
		\vspace{-2em}
		\includegraphics[width = 0.35\textwidth]{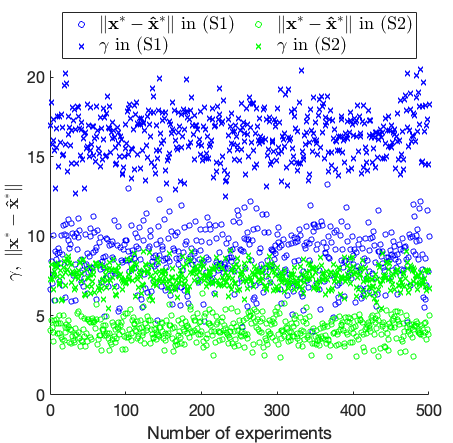}
		\caption{The comparison of $\gamma$ and $\|\bfx^*-\hat{\bfx}^*\|$ between two parameter configurations (S1) and (S2).}
		\label{fig:gamma}
		\end{figure}
	
		In Fig.~\ref{fig:gamma}, we plot the comparison of $\gamma$ and $\|\bfx^*-\hat{\bfx}^*\|$ between two parameter configurations (S1) and (S2). The result of Fig.~\ref{fig:gamma} is consistent with Theorem~\ref{thm:gamma_bound} that the distance between the original NE and the perturbed NE is bounded by $\gamma$ while preserving differential privacy. 
	 \hfill$\square$
	\end{experiment}

	\begin{experiment}[Benchmark with Existing Methods]
	\rm  Based on Experiment~\ref{exp:validations}, we further plot each player's original NE and distribution of the perturbed NE under two parameter configurations (S1) and (S2) among $500$ executions.
	
 	The result of Fig.~\ref{fig:distribution_ne} shows that most perturbed NE under (S1) and (S2) are located to the left of the original NE. The parameter configuration (S2) has a weaker requirement for differential privacy. The perturbed NE under (S2) are closer to the original NE, implying that one has to sacrifice the differential privacy of payoff functions for  the accuracy of NE. 
 	
 	\begin{figure}[tb]
 	\centering
 	\includegraphics[width = 0.35\textwidth]{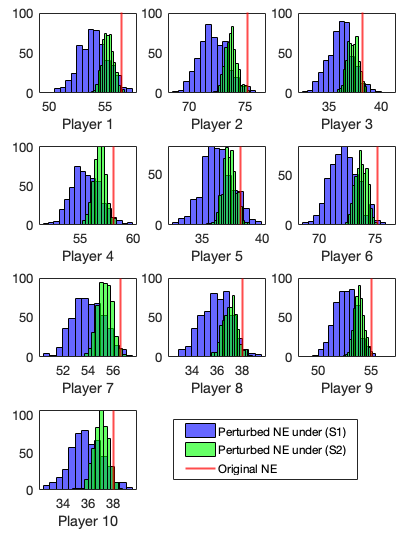}
 	\caption{The original NE and the distribution of the perturbed NE under the parameter configurations (S1) and (S2). The y-axis represents the number of times that player $i$'s perturbed NE action occurred within the intervals set by the x-axis among $500$ executions.}
 	\vspace{-1.5em}
 	\label{fig:distribution_ne}
	 \end{figure}
 
 	To further show the relevance among the accuracy of NE, the privacy requirement $\epsilon$ and the Laplace parameter $a$, we fix $\delta = 0$ and  plot $\|\bfx^*-\hat{\bfx}^*\|/\|\bfx^{\ast}\|$ versus $\epsilon$ and $a$ each for $50$ executions in Fig.~\ref{fig:distance_vs_a}. Roughly speaking, it shows that as $\epsilon$ decreases (stricter privacy), the Laplace parameter $a$ increases and thus the accuracy of NE decreases. 
 	
 	Compared with existing methods of state/communication perturbation\cite{ding2021differentially,han2016differentially,ye2021differentially}, we both show the tradeoff between the accuracy of the optimal points/NE and the privacy of objective functions/payoff functions, However, they prove that $\lim_{k \to \infty} \mathbb{E}(\| \bfx(k) - \bfx^{\ast}\|^{2})$ has a upper bound depending on convergence rate and privacy level. According to the structure of their privacy algorithm, they might produce a asymptotically unbiased perturbed NE. Unlike those results, Algorithm~\ref{alg:laplace_lq} always produces a biased perturbed NE. It arises from $\mathbb{E}(q_{ii})>0,\forall i \in \rmV$ that are necessary to ensure the concavity of the perturbed payoff function. However, they have to add perturbation to communications at all time steps. For example, in \cite{ding2021differentially}, it takes roughly $4000$ steps to reach a close distance from $\bfx^{\ast}$, in each of which perturbation is added to players' state. In contrast, we only add perturbation to the original payoff functions once, after which the computation of the distributed NE is deterministic. The number of linear-quadratic perturbation coefficients (non-zero $q_{ij},\beta_{i},\forall i,j \in \rmV$) generated by Algorithm~\ref{alg:laplace_lq} is only $60$, which is far less than theirs.   	\hfill $\square$
 	\begin{figure}[tb]
 		\centering
 		\includegraphics[width = 0.35\textwidth]{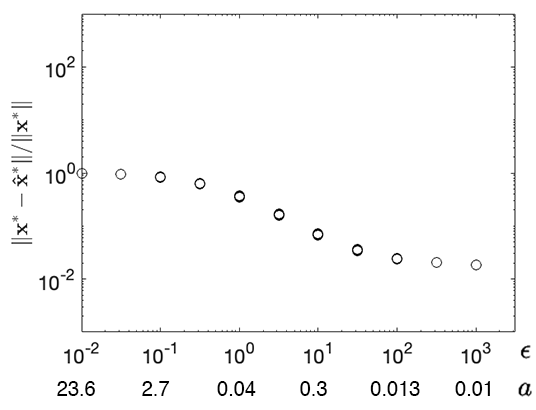}
 		\caption{The plot of $\|\bfx^*-\hat{\bfx}^*\| / \| \bfx^{\ast}\| $ versus $\epsilon$ and $a$ each for $50$ executions.}
 		\label{fig:distance_vs_a}
 		\vspace{-1.5em}
 	\end{figure}
 
	\end{experiment}

	\begin{experiment}[Tradeoff between Privacy and Payoffs]
		\rm  Upon Experiment~\ref{exp:validations}, we further compute the players' payoffs at the original NE and the perturbed NE under two parameter configurations, $\bff(\bfx^{\ast})$ and $\bff(\hat{\bfx}^{\ast})$, and plot them in Fig.~\ref{fig:payoffs},. 
		
		From the result of Fig.~\ref{fig:payoffs}, it is not surprising that players' payoffs at the perturbed NE under (S1) and (S2) are always lower than those at the original NE.  Players' payoffs at the perturbed under (S1) are lower than those at the perturbed NE under (S2). It indicates that the sacrifice of the accuracy of NE for payoff functions' privacy leads to the decline of players' payoffs.
		\begin{figure}[h]
		\centering
		\vspace{-1em}
		\includegraphics[width = 0.35\textwidth]{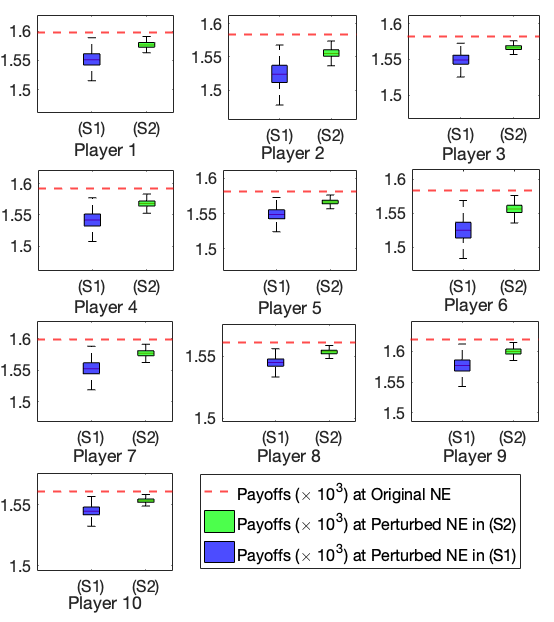}
		\caption{Players' payoffs at the original NE, the perturbed NE in (S1), and the perturbed NE in(S2). The y-axis represents the value of player $i$'s payoffs.}
		\label{fig:payoffs}
		\vspace{-1.5em}
		\end{figure}
	 \hfill$\square$
	\end{experiment}

	\section{Conclusion}\label{sec:conclusion}
	In this work, we investigated network games in which players participated in information aggregation processes under the differential privacy requirement for players' payoff functions. The LLQFP mechanism was proposed. We turned to monotone games, demonstrating that the LLQFP mechanism preserved the concavity property and generated a bounded perturbed NE which was controllable by Laplace parameter tuning. We also looked at LQ games as a pedagogical example to explain, given what Laplace parameter conditions, differential privacy of the LLQFP mechanism could be ensured.
	Finally, numerical examples were presented to demonstrate the benefits of the LLQFP mechanism. 
\bibliographystyle{IEEEbib}
\bibliography{reference}

\end{document}